\documentclass[12pt]{article}

\usepackage{graphicx}

\usepackage{amssymb}
\usepackage{amsmath}

\begin{document}
	
\title{Testing a hydroacoustic radiator in a reverberant tank based on recording the sound field in the air above the tank}
\author{A.L. Virovlyansky, M.S. Deryabin, A.A. Prokhorov,\\ A.Yu. Kazarova, V.K. Bakhtin\\
	\textit{{\small A.V. Gaponov-Grekhov Institute of Applied Physics of the Russian Academy of Sciences}}\\
	\textit{{\small 46 Ul'yanov Street, 603950 Nizhny Novgorod,
			Russia}} \\
	\textit{{\small virovlyansky@mail.ru}}\\
	\date{}} \maketitle

\begin{abstract}
A method for calibrating a monopole sound source in a water tank with reflective side walls and bottom is considered. The idea of the method is based on the phenomenon of anomalous transparency of the water-air boundary for a sound source located at a shallow depth. This boundary plays the role of a filter that prevents waves reflected from the side walls and bottom from entering the air. For a shallow source, the field in the air will be approximately the same as for a source located at the same depth in a homogeneous water half-space. This field is described by a well-known analytical formula that makes it possible to estimate the source strength  in water based on the sound intensity level measured in air.
\end{abstract}

\section{Introduction}

Traditional approaches to solving the problem of free-field calibration of the acoustic radiator in a reverberant tank are based on the isolation of the direct signals arriving at the receivers without reflection from the tank boundaries. There are two conventional methods for eliminating the effect of boundary echoes \cite{Bobber,Robinson1999}. One of them is to coat the boundaries with absorbers. The second method relies on the time-gating technique for resolving the direct signals by their arrival times. However, both methods are applicable only at sufficiently high frequencies. There exist more sophisticated methods which can be used at acoustic frequencies below the limits imposed by the echo-free time of the test tank \cite{Robinson2018,Isaev2017,2E-2U,wang2025three}. A calibration method that does not require the direct signal contribution to be isolated is proposed in Refs. \cite{V2019,V2020b}. It is based on comparing the field generated in the tank by the source to be calibrated with the fields generated in the same tank by a reference acoustic monopole.

This paper considers a method for calibrating a monopole source based on measurements of the sound field that this source emits from the tank into the air. It is well known that the interface between water and air half-spaces for plane waves is almost completely reflective \cite{Chapman1990}. However, for a source located in water at a shallow depth, the effect of anomalous transparency of the boundary occurs, as discovered in Ref.
\cite{Godin2006} (see also \cite{Godin2007_eng,Godin2008,Godin2013_eng}). It is assumed that the size of the source is small compared to the wavelength and the field excited by it represents a superposition of homogeneous and inhomogeneous (evanescent) plane waves. Near the source, the latter dominate, but their amplitudes decrease exponentially with distance. If the source depth is small compared to the wavelength, the amplitudes of a significant part of the inhomogeneous waves on the way to the boundary do not have time to weaken noticeably, and after crossing the boundary these waves become propagating. Then 'the ratio of energy radiated into the air to the total emitted
energy is larger by a factor of up to 3400 ... than it is for the same source
located a wavelength or more from the interface' \cite{Godin2008}.

A similar effect may be observed for a source in a water tank. We consider a situation where the sound field in the air above a shallow acoustic monopole in a tank is approximately described by a well-known analytical expression for the field above a monopole in a water half-space \cite{Godin2006,Godin2007_eng,Godin2008}. Using this relation allows one to estimate the sound pressure level (SPL) of a source in water by measuring the sound field in the air. This is the idea of our approach.

We assume that the water-air boundary transmits those excited shallow source waves that propagate toward the water surface, but leaves in the tank the waves reflected from the bottom and side walls. Analyzing the conditions for the applicability of this assumption is a complex task, the solution of which is beyond the scope of this work. Here, we limit ourselves to considering a specific example that demonstrates the effectiveness of the method.

The paper is organized as follows. In Sec. \ref{sec:water}, following Ref. \cite{Godin2007_eng}, we derive an analytical expression for the sound field in air excited by an acoustic monopole in a water half-space. In Sec. \ref{sec:tank}, using finite element modeling (FEM) in COMSOL Multiphysics, it is shown that this expression can correctly describe the field in the air above the tank, excited by an oscillating sphere of small radius located in the tank at a shallow depth. Section \ref{sec:test} describes a feasibility test that demonstrates the applicability of the method for estimating the SPL of a monopole sound source. The results of the paper are summarized in Sec. \ref{sec:Conclusion}.

\section{Sound transmission from water to air \label{sec:water}}

Consider a situation where a point sound source located in a water half-space excites a sound wave in the air half-space above the water. Let us introduce a cylindrical coordinate system $\left( r,\varphi,z\right)  $ with a vertical axis $z$ pointing upwards. We assume that the horizontal plane $z=0$ coincides with the  water -- air boundary. We consider the air ($z>0$) and water ($z<0$) half-spaces to be homogeneous with densities $\rho_{a}=1.012$ kg/m$^{3}$ and $\rho_{w}=1000$ kg/m$^{3}$ and velocities $c_{a}=343$ m/s and $c _{w}=1470$ m/s, respectively. A source located at a point with coordinates $r=0$ and $z=z_{0}$, where $z_{0}<0$, excites a sound field at an angular frequency $\omega$. The field does not depend on the azimuthal angle $\varphi$.

To make this paper self-explanatory, we present a brief derivation of the explicit expression for the complex amplitude of the excited field in air. The following notations will be used:%

\[
k_{a}=\omega/c_{a}\text{,\ }k_{w}=\omega/c_{w}\text{,\ }\gamma_{a}\left(
q\right)  =\sqrt{k_{a}^{2}-q^{2}}\text{, \ }\gamma_{w}\left(  q\right)
=\sqrt{k_{w}^{2}-q^{2}}.
\]

We assume that the source is an acoustic monopole that excites a field in the water \cite{BG99}%
\[
p_{i}\left(  r,z\right)  =\frac{1}{\sqrt{r^{2}+\left(  z-z_{0}\right)  ^{2}}%
}\mathrm{e}^{\mathrm{i}k_{w}\sqrt{r^{2}+\left(  z-z_{0}\right)  ^{2}}}%
\]%
\begin{equation}
=\mathrm{i}\int_{0}^{\infty}\frac{q\mathrm{d}q}{\gamma_{w}\left(  q\right)  }J_{0}\left(
qr\right)  \mathrm{e}^{\mathrm{i}\gamma_{w}\left(  q\right)  \left\vert z-z_{0}\right\vert },
\label{p-i}%
\end{equation}
where $J_{0}\left(  qr\right)  $ is the Bessel function of the first kind of order zero. Here and below, we omit the time factor $\mathrm{e}^{-\mathrm{i}\omega t}$. The function $p_{i}\left(  r,z\right)  $ describes the field incident on the water-air interface. The reflected field is expressed by the integral%
\begin{equation}
p_{r}\left(  r,z\right)  =\mathrm{i}\int_{0}^{\infty}\frac{q\mathrm{d}q}{\gamma_{w}\left(
q\right)  }V\left(  q\right)  J_{0}\left(  qr\right)  \mathrm{e}^{-\mathrm{i}\gamma_{w}\left(
q\right)  z}, \label{p-r}%
\end{equation}
where $V\left(  q\right)  $ is the reflection coefficient of the component with horizontal wave number $q$. Thus, the total field in the region $z<0$ (in water)%
\begin{equation}
p\left(  r,z\right)  =p_{i}\left(  r,z\right)  +p_{r}\left(  r,z\right)  .
\label{p-w}%
\end{equation}
In the region $z>0$ (in air), the sound field is equal to%
\begin{equation}
p\left(  r,z\right)  =\mathrm{i}\int_{0}^{\infty}\frac{q\mathrm{d}q}{\gamma_{a}\left(  q\right)
}W\left(  q\right)  J_{0}\left(  qr\right)  \mathrm{e}^{\mathrm{i}\gamma_{a}\left(  q\right)
z}, \label{p-a}%
\end{equation}
where $W\left(  q\right)  $ is the transmission coefficient.

The coefficients $V\left( q\right) $ and $W\left( q\right) $ are found from the relations \cite{BL2003} $p_{w}\left(  r,0\right)  =p_{a}\left(  r,0\right)  $ and
\[
\frac{1}{\rho_{w}}\left.  \frac{\partial p\left(  r,z\right)  }{\partial
z}\right\vert _{z=-0}=\frac{1}{\rho_{a}}\left.  \frac{\partial p\left(
r,z\right)  }{\partial z}\right\vert _{z=+0},
\]
expressing the continuity of pressure and normal component of velocity at the boundary $z=0$, respectively. Substituting (\ref{p-i}) -- (\ref{p-a}) here and equating the left and right sides of the integrals for the same $q$, we find $V\left(  q\right)  $ and $W\left(  q\right)  $. The expression for the transmission coefficient is%

\[
W\left(  q\right)  =\frac{2m\gamma_{a}\left(  q\right)  }{m\gamma_{w}\left(
q\right)  +\gamma_{a}\left(  q\right)  }\mathrm{e}^{\mathrm{i}\gamma_{w}\left(  q\right)  z_{0}%
}.
\]
Substituting it into (\ref{p-a}), we find%
\begin{equation}
p\left(  r,z\right)  =2\mathrm{i}m\int_{0}^{\infty}\frac{q\mathrm{d}q}{m\gamma_{w}\left(
q\right)  +\gamma_{a}\left(  q\right)  }J_{0}\left(  qr\right)  \mathrm{e}^{\mathrm{i}\gamma
_{a}\left(  q\right)  z-i\gamma_{w}\left(  q\right)  z_{0}}\text{, \ }z>0.
\label{pa}%
\end{equation}

Note that the components of the field excited in water (\ref{p-i}), corresponding to wave numbers $q<k_{w}$, describe homogeneous (propagating) waves, while components with $q>k_{w}$ describe inhomogeneous waves, which decay exponentially with distance from the source. After crossing the water-air interface, inhomogeneous waves with $q$ from the interval $k_{w}<q<k_{a}$ become propagating. Waves with $q>k_{a}$ do not propagate in air. The attenuation of an inhomogeneous wave on its way from the source to the interface is determined by the factor $\exp\left(  -\left\vert k_{w}^{2}-q^{2}\right\vert ^{1/2}|z_{0}|\right)  $ and is greater the deeper the source is. For $|z_{0}|>\lambda_{w}$, where $\lambda_{w}=2\pi /k_{w}$ is the wavelength in water, the amplitudes of inhomogeneous waves at the interface are negligible. The effective integration interval in (\ref{pa}) extends from 0 to $k_{w}$. As the depth of the source decreases, inhomogeneous waves appear, which do not have time to attenuate significantly and make an additional contribution to the sound field in the air. When the condition $\left(  k_{a}^{2}-k_{w}^{2}\right)  ^{1/2}|z_{0}|\ll1$ is satisfied, the upper limit of the effective integration interval in (\ref{pa}) increases to the maximum possible value $k_{a}$.
Considering that $k_{a}\simeq4.3$ $k_{w}$, rewrite the above inequality as%
\begin{equation}
|z_{0}|\ll\lambda_{a}, \label{z-la}%
\end{equation}
where $\lambda_{a}=2\pi/k_{a}$ is the wavelength in air. At such small $|z_{0}|$, the sound field in air is significantly amplified by contributions from inhomogeneous waves
\cite{Godin2006,Godin2007_eng,Godin2008,Godin2013_eng}.

\section{Numerical simulation of sound radiation from a tank into the air
\label{sec:tank}}

This section presents a numerical example demonstrating that, under certain conditions, the expression (\ref{pa}) can correctly describe the field in air generated by a small-radius monopole source located in a tank.

Figure 1 shows the results of calculating the amplitude of sound pressure on the water surface created by a point source located in the water half-space at points $(0,z_0)$ with $z_0$ = - 3 cm (a) and $z_0$ = - 7 cm (b). Here and below, we consider a point source that excites a field (\ref{p-i}) in a homogeneous water space. The amplitude of this field at a distance of 1 m from the source is 1 Pa. The calculation is performed at frequencies of 5 kHz (a) and 20 kHz (b). As we see, at both frequencies, the sound field at a distance of 0.6 m from the source's projection on the water surface is significantly attenuated. This suggests that the field at frequencies of 5-20 kHz, excited in air by a shallow monopole source from a tank whose walls are at least 0.6 m from the source, may be approximately described by Eq. (\ref{pa}).

\begin{figure}[!htbp]
	\centering
	\includegraphics[width=4.5in]{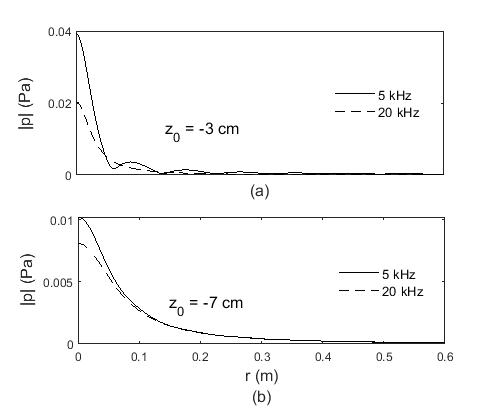}\caption{Pressure magnitude on the water surface at frequencies of 5 kHz (solid line) and 20 kHz (dashed line). The point source exciting the sound field is located at depths of -3 cm (a) and -7 cm (b).}%
	\label{fig1}%
\end{figure}

To verify this assumption, we performed FEM of sound excitation in air by a monopole source from a water-filled tank shaped like a cylinder 1 m high and 0.6 m in radius. A pulsating sphere was considered as the source. The center of the sphere is on the tank axis, and therefore the sound field is axially symmetric. The axially symmetric numerical model includes a water cylinder and an air cylinder of the same radius and height of 0.3 m located above it (Fig. 2). The air cylinder is surrounded by perfectly matched layers, and two boundary conditions are considered for the side surface and bottom of the water cylinder: (i) a hard boundary, where the normal component of velocity is zero, and (ii) a soft boundary, where the pressure is zero.

\begin{figure}[!htbp]
	\centering
	\includegraphics[width=4.5in]{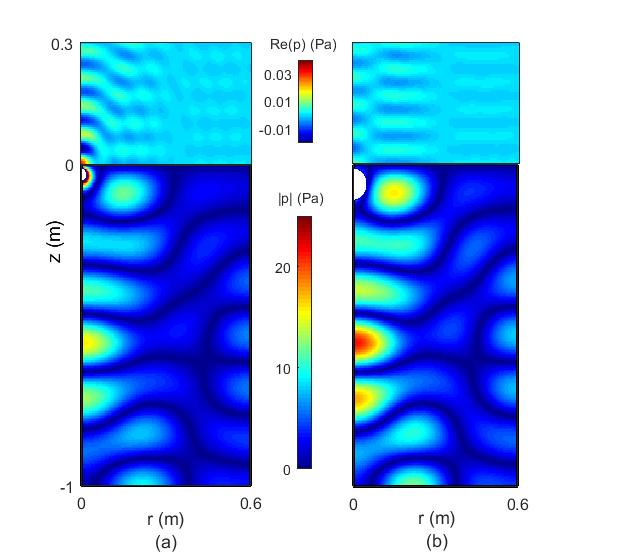}\caption{Pressure magnitude in a cylindrical water tank with a solid side boundary and bottom (-1 m $<$ $z$ $<$ 0 m) and the real part of the complex pressure amplitude in the air above the tank (0 m $<$ $z$ $<$ 0.3 m). The sound source is a sphere with a radius of 2 cm (a) and 5 cm (b), pulsating at a frequency of 5150 Hz. The center of the sphere is at a depth of -3 cm (a) and -6 cm (b).}%
	\label{fig2}%
\end{figure}

Calculations were performed for spheres with radii $a$ = 2 cm and $a$ = 5 cm. In both cases, the volume velocities of the pulsating sphere were chosen to be equal to the volume velocity of the point source.

Figure 2 shows the distributions of the real part of the complex pressure amplitude $\text{Re}(p)$ in air and the pressure magnitude $|p|$ in water for a tank with a solid side surface and bottom. In air, the field is formed by propagating waves. The magnitudes of the real and imaginary parts of their pressure are of the same order. In a tank (resonator), where the field is formed predominantly by standing waves, the ratio of the magnitudes of the real and imaginary parts of the pressure is determined by the phase of the source complex amplitude. This ratio can be arbitrarily large or small (in our case, the imaginary part dominates). The distribution of $|p|$ shown in the figure does not depend on the phase of the source amplitude. As we can see, the pressure in water is 1000 times greater than in air. The calculation results presented in Fig. 2 were obtained for spheres with a radius of 2 cm (a) and 5 cm (b) pulsating at a frequency of 5150 Hz. The centers of the spheres are located at depths of -3 cm and -6 cm, respectively. 

In Fig. 3, we compare the depth dependence of the air pressure amplitude at a frequency of 5150 Hz, predicted by the relation (\ref{pa}) (the same bold solid line on all graphs) and the results of finite element calculations obtained for a tank with a solid (thin solid lines) and soft (dashed lines) side surface and bottom. The calculations were performed for 4 pairs of values ($a$, $z_{0}$): (2 cm,-3 cm) (a), (5 cm,-6 cm) (b), (2 cm,-7 cm) (c), and (5 cm,-10 cm) (d). Plots (a) and (b) ((c) and (d)) are constructed for spheres with upper points at a depth of -1 cm (- 5 cm), respectively. Numerical simulation shows that Eq. (\ref{pa}) describes the pressure in the air relatively well only for a sphere of small radius, the center of which is at a depth of $z_{0}$ = -3 cm (Fig. 3(a)). The influence of the  size of the pulsating sphere on the excitation of the field in air is analyzed in Ref. \cite{Godin2013_eng}.   

\begin{figure}[!htbp]
	\centering
	\includegraphics[width=4.5in]{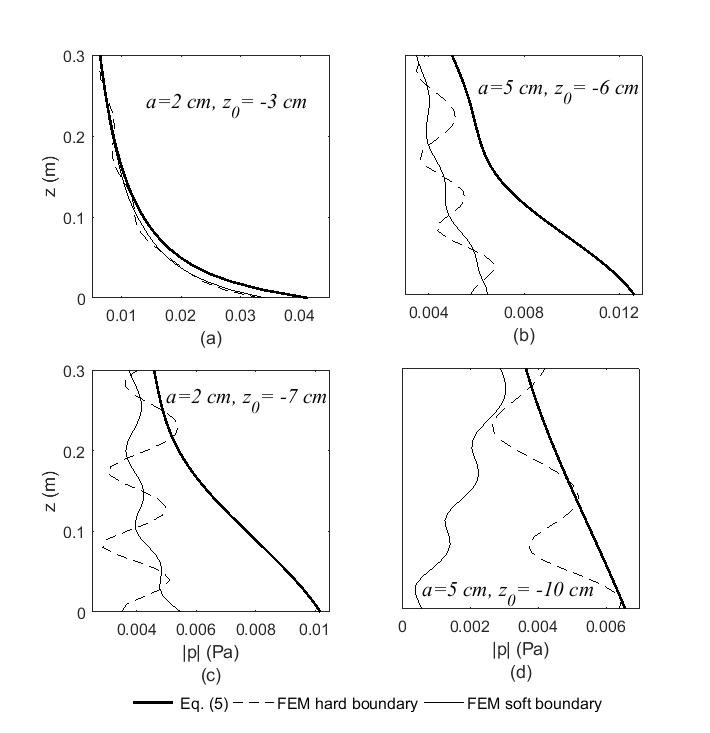}\caption{Air pressure at $r = 0$ generated by a source in a water half-space (solid thick line) and in a tank with rigid (thin solid line) and soft (dashed line) side boundaries and bottom.The calculations are performed for sources whose radii, $a$, and center depths, $z_0$, are indicated on the graphs.}%
	\label{fig_3}%
\end{figure}

Note that Eq. (\ref{pa}) becomes inapplicable as $f$ approaches the resonant frequency of the water tank. The frequency of 5150 Hz, at which the finite element modeling was performed, was chosen such that it lies approximately midway between adjacent resonant frequencies of the tank with a soft upper boundary. This holds for both boundary conditions on the side surface and bottom. Figure 4 (a) shows the same field calculation results as in Fig. 2 (a), but performed at a frequency of $f$ = 5110 Hz. This value is close to one of the resonant frequencies of our tank with a solid side boundary and bottom equal to 5108 Hz. The field in the water is determined by the dominant mode and differs significantly from the field of a point source in the water half-space.Accordingly, the dependence of the amplitude of the air pressure above the source on the height of the receiving point, shown in Fig. 4 (b) by the dashed line, differs significantly from the similar dependence calculated using formula (\ref{pa}) and represented in Fig. 4(b) by the same solid line as in Fig. 3 (a).

\begin{figure}[!htbp]
	\centering
	\includegraphics[width=4.5in]{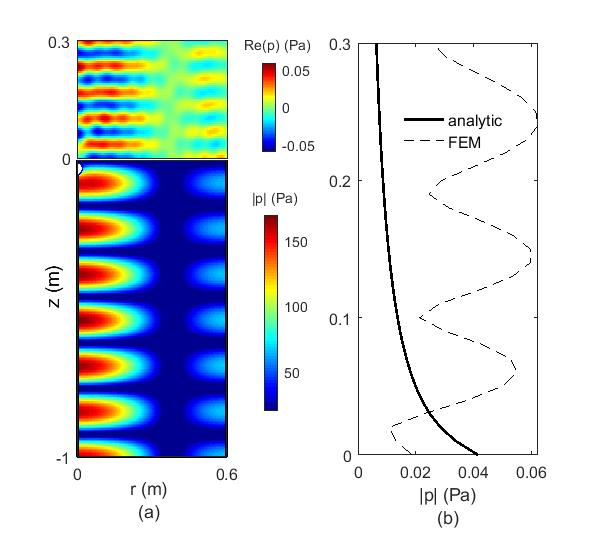}\caption{(a) The same as in Fig. 2(a) but for frequency $f$ = 5110 Hz. (b) The same as in Fig. 3(a) but for frequency $f$ = 5110 Hz.}%
	\label{fig4}%
\end{figure}

\section{Feasibility test \label{sec:test}}

This section describes the results of applying the discussed approach to free field calibration of monopole acoustic sources in the frequency range of 5-20 kHz. Two reversible hydrophones of the B\&K 8100 and B\&K 8103 types were calibrated (Fig. 5). In a homogeneous water space, each of these sources at frequencies in the specified range excites a field that is approximately described by the expression $p_{0}\exp\left( \mathrm{i}k_{w}R\right) /R$, where $R$ is the distance to the observation point, and $p_{0}$ specifies the pressure at a distance of 1 m from the source. The task of calibration was to estimate $\left\vert p_{0}\right\vert $, corresponding to the amplitude of a voltage of 1 V applied to the source.

\begin{figure}[!htbp]
	\centering
	\includegraphics[width=4.5in]{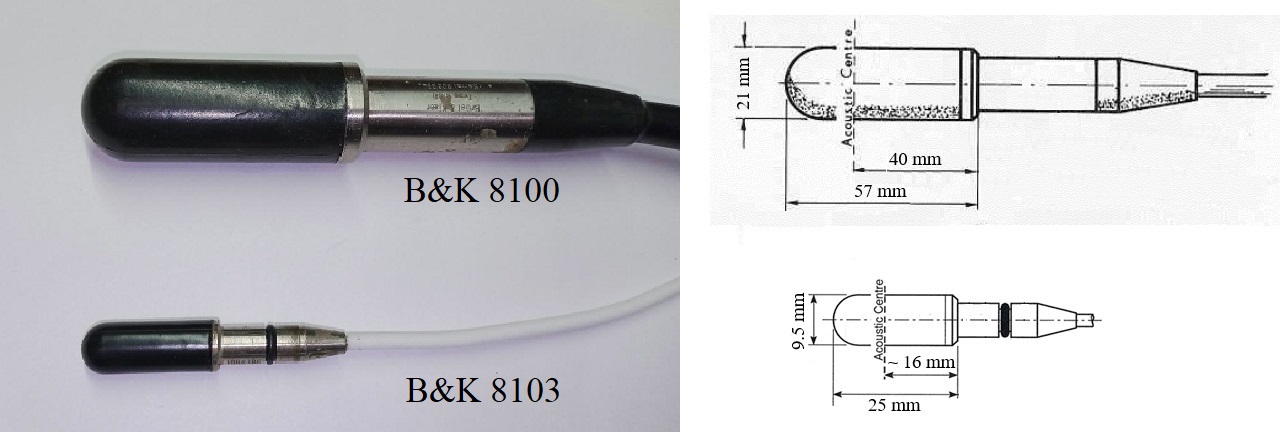}\caption{Calibrated sources (left) and schematic drawings of their constructions (right).}%
	\label{fig_5}%
\end{figure}

The tank in which the measurements were taken was a cube measuring $1\times1\times1$ m with a free surface and side walls and bottom made of organic glass.

\begin{figure}[!htbp]
	\centering
	\includegraphics[width=4.5in]{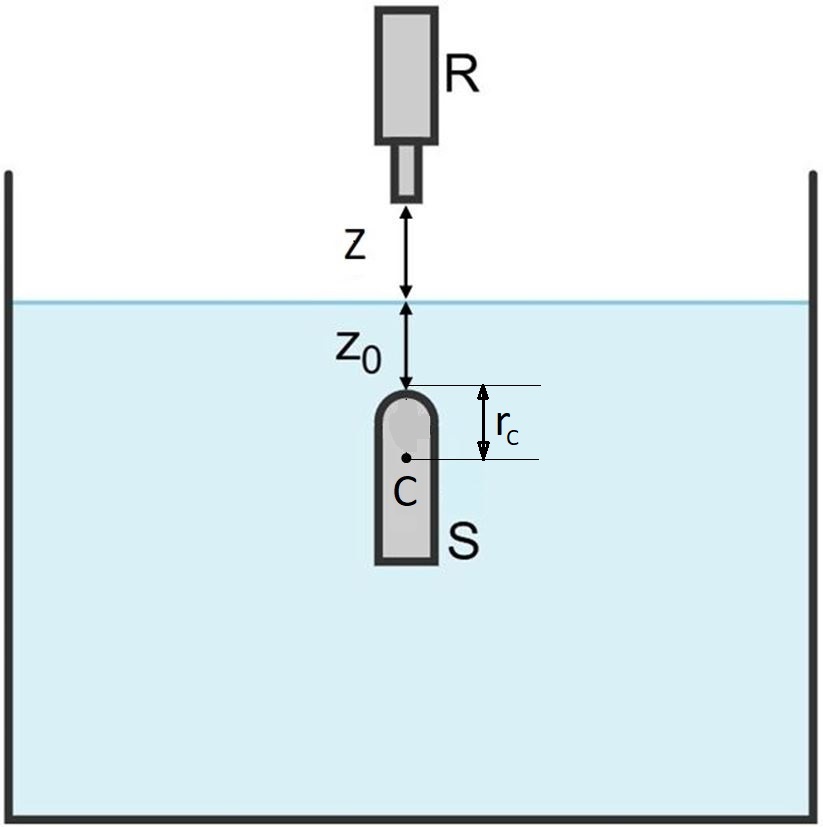}\caption{Measurement setup.}%
	\label{fig_6}%
\end{figure}

The measurement setup is shown in Fig. 6. Source $S$, located in water at a depth of $z_{0}$, emits a tonal signal at a frequency $f$, which is recorded in the air at a height $z$ by microphone R. The acoustic center of the source is located at point $C$, whose depth is equal to $z_{0}+r_{c}$. Let $P_{m}\left(  f\right)  $ be the amplitude of the pressure in the air at a height $z$ above the source recorded by the microphone, and $P\left(  f\right)  $ is the amplitude of the pressure at frequency $f$ at point $\left(  0,z\right)  $, calculated using formula (\ref{pa}) for a point source at a depth of $z_{s}=z_{0}+r_{c}$.
According to our assumption made in the previous section, when
$z_{0}$ is small, the desired estimate $\left\vert p_{0}\right\vert $ has the form

\begin{equation}
\left| p_{0} \right| = \left.  P\left( f\right) \right/ P_{m}\left(  f\right).
\label{p0}%
\end{equation}

Both sources were placed at a depth of $z_{0}$ = 1 cm. The calculations used the values of $r_{c}$ specified in the documentation of the manufacturer \cite{BK}: 1.7 cm for B\&K 8100 and 0.9 cm for B\&K 8103 (see Fig. 3).

In Fig. 7 and 8, the thin solid lines show the frequency dependences of%

\begin{equation}
\text{SPL}=20\log_{10}\left(  \frac{\left\vert p_{0}\right\vert }%
{10^{-6}\text{Pa}}\right)  \text{,}\label{SPL}%
\end{equation}
found using the relation (\ref{p0}) for sources B\&K 8100 and B\&K 8103, respectively. The resulting SPL frequency dependence does not exhibit any significant outliers. This indicates that the modes in our tank have low Q factors, and resonant effects do not limit the capabilities of the method.

\begin{figure}[!htbp]
	\centering
	\includegraphics[width=4.5in]{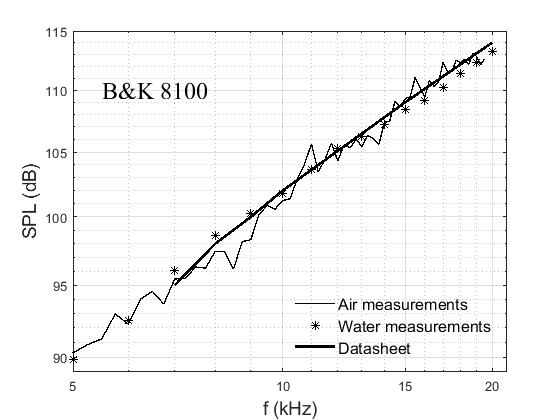}\caption{SPL values for the B\&K 8100 source, obtained from measurements in air (thin solid line), measurements in water (asterisks) and taken from the manufacturer's official datasheet (thick solid line).}%
	\label{fig_7}%
\end{figure}

\begin{figure}[!htbp]
	\centering
	\includegraphics[width=4.5in]{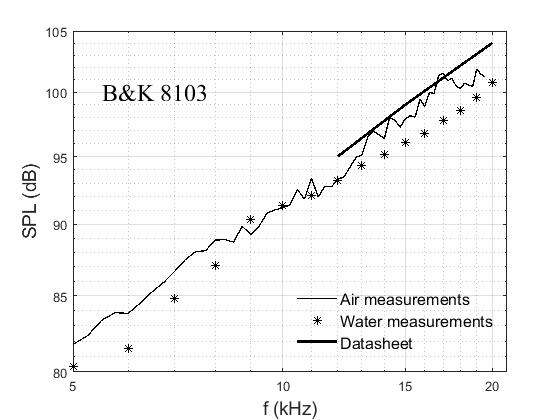}\caption{The same as in Fig. 7, but for source B\&K 8103.}%
	\label{fig_8}%
\end{figure}

The asterisks in Figs. 7 and 8 indicate the results of an alternative SPL estimate at frequencies of 5, 6,..., 20 kHz, based on measurements of the source field in water. Let us dwell on this in more detail.

The measurement setup is shown in Fig. 9. Source S emits a signal $s_{0}$ consisting of five periods of carrier frequency $f$ (Fig. 10). It is recorded by receiving hydrophone $R$ located at a distance $r$ from the edge of the source. The source and receiver are located approximately in the center of the tank, with the distance $r$ varying from 5 to 50 mm in 5 mm increments. Fig. 10 shows examples of the emitted (a) and received (b) signals. To estimate $\left\vert p_{0}\right\vert $, the dependence of the value of the third maximum of the signal, $p_{3}$, marked with an arrow, on the distance $r$ is used. It is assumed that

\[
\left\vert p_{3}\right\vert =\frac{\left\vert p_{0}\right\vert }{r+r_{c}}.
\]

The desired estimate is given by the value $\left\vert p_{0}\right\vert $, for which the function on the right-hand side best approximates the sequence of measured values $\left\vert p_{3}\right\vert $.

\begin{figure}[!htbp]
	\centering
	\includegraphics[width=4.5in]{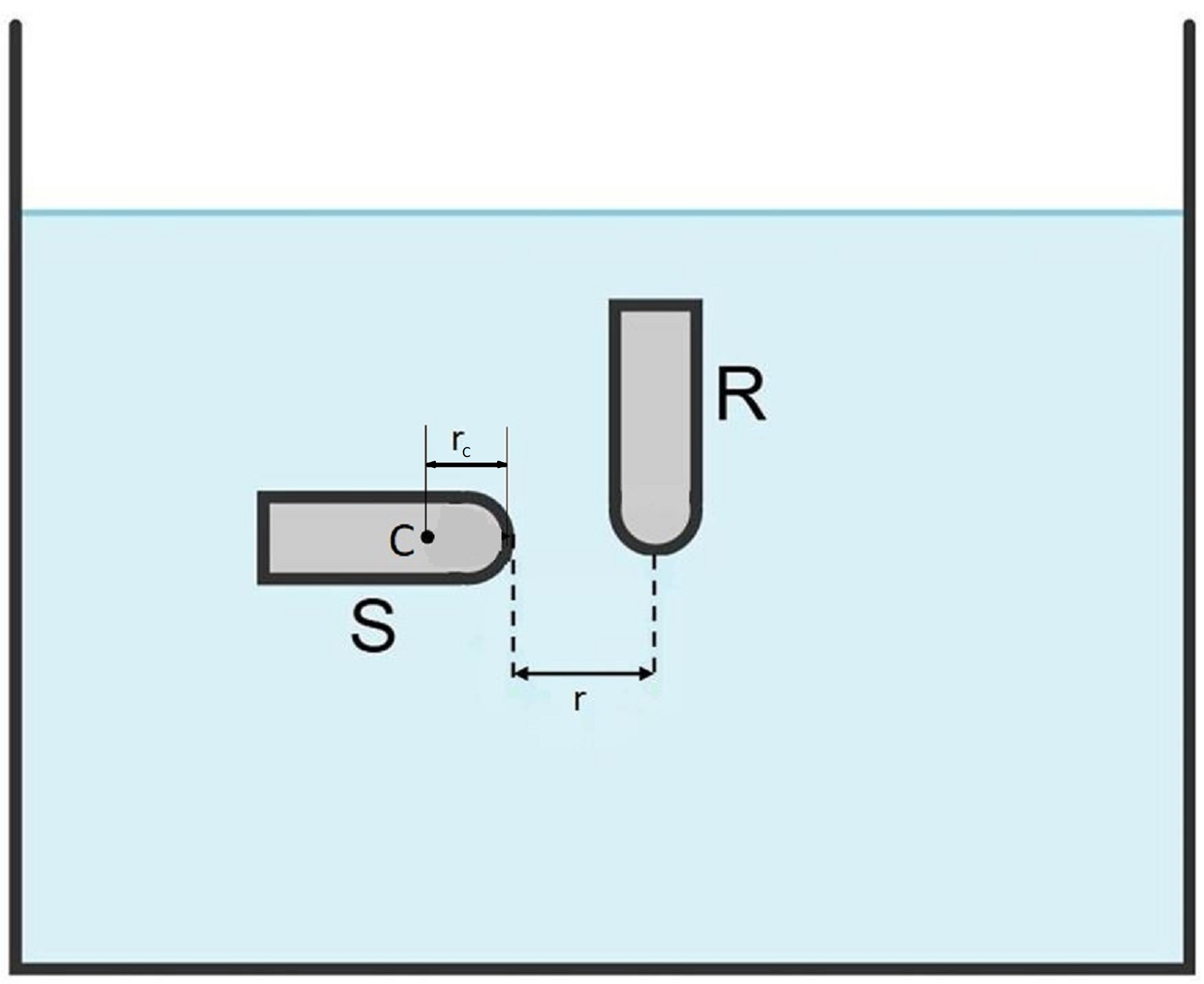}\caption{Experimental setup for alternative SPL estimation based on tank sound field measurements.}%
	\label{fig_9}%
\end{figure}

\begin{figure}[!htbp]
	\centering
	\includegraphics[width=4.5in]{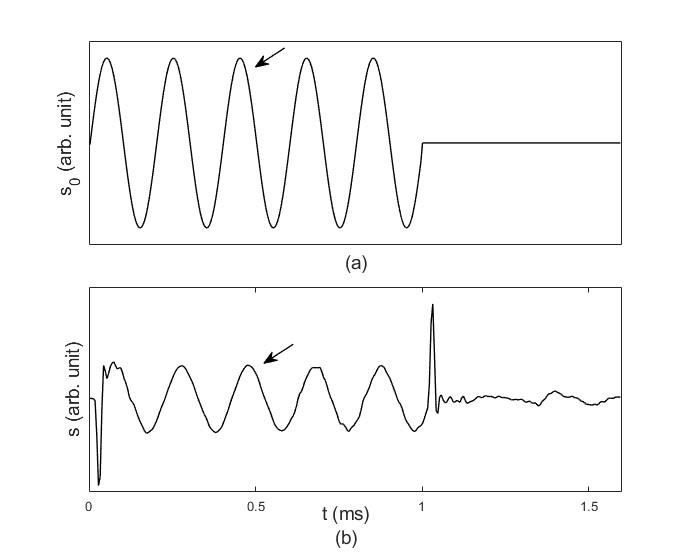}\caption{Examples of transmitted (a) and time-gated received (b) signals from five carrier frequency periods. The arrow marks the third signal maximum, the amplitude of which is used to estimate the SPL.}%
	\label{fig_10}%
\end{figure}

The bold solid lines in Figs. 7 and 8 represent SPL values taken from the manufacturer's official datasheet \cite{BK}.

\section{Conslusion \label{sec:Conclusion}}

The comparison results of SPL estimates obtained by different methods, presented in Figures 7 and 8, demonstrate the effectiveness of the approach under consideration. They confirm that the water-air interface can indeed act as a filter, allowing waves emitted from a shallow source to pass upwards into the air, that is, toward the receiver, and retaining waves reflected from the walls and bottom within the tank. In fact, this interface plays the same role as the absorbing surface coating of the tank or time-gating during calibration based on sound field measurements in water.

However, the conditions for applying this approach require further study. Requirements for the size of the tank, the size of the source, and its depth need to be formulated. In addition, the position of the acoustic center of the source must be known. In cases where it does not coincide with the geometric center, this constitutes a separate task. It should also be borne in mind that the SPL of a monopole source is determined by its volume velocity. The value of this velocity, and with it the SPL, for a source oscillating near the water surface may differ from its values at greater depths.

To apply the sound source calibration method under consideration, it is necessary to have a relationship linking the source's SPL and the intensity of the sound field recorded in the air. In our work, this relationship is given by Eq. (\ref{pa}), which defines the sound field excited by a point source from an infinite water half-space. It can be used when calibrating a monopole source.  The works \cite{Godin2006,Godin2007_eng} provide formulas that generalize Eq. (\ref{pa}) for a wide class of sources. In particular, these works provide an analogue of Eq. (\ref{pa}) for a dipole source. In principle, these relations can be used when calibrating sources that are not monopoles.  

We assume that FEM can be used to establish a connection between the SPL of the calibrated source and the field in air. We also assume that the possibilities for calibrating the source based on measurements in air can be expanded using the capabilities provided by metasurfaces, which allow changing the impedance of the water-air boundary \cite{ms24,ms25}. However, questions about calibrating a non-monopole sound source and the possibility of using metasurfaces are beyond the scope of this work.

\section*{CRediT authorship contribution statement}

{\bf A.L. Virovlyansky}: Writing -- original draft, Conceptualization, Methodology, Formal analysis. {\bf M.S. Deryabin}: Investigation, Data curation, Formal Analysis. {\bf A.A. Prokhorov}: Investigation, Data curation. {\bf A.Yu. Kazaarova}: Writing -- review end editing, Methodology, Software. {\bf V.K. Bakhtin}: Methodology, Software.

\section*{Declaration of competing interest}

The authors declare that they have no known competing financial interests or personal relationships that could have appeared to influence the work reported in this paper.

\section*{Acknowledgements}

This work was performed within the State assignment for the Institute of Applied Physics of the Russian Academy of Sciences (project FFUF-2024-0041).

\end{document}